\documentclass[aps,preprint]{revtex4}
\usepackage{amsfonts}
\usepackage{amsmath}
\usepackage{amssymb}
\usepackage{graphicx}

\input{tcilatex}

\begin{document}

\preprint{}
\title[ ]{Lovelock black holes with a power-Yang-Mills source}
\author{S. Habib Mazharimousavi}
\email{habib.mazhari@emu.edu.tr}
\affiliation{Department of Physics, Eastern Mediterranean University, G. Magusa, north
Cyprus, Mersin 10, Turkey.}
\author{M. Halilsoy}
\email{mustafa.halilsoy@emu.edu.tr}
\affiliation{Department of Physics, Eastern Mediterranean University, G. Magusa, north
Cyprus, Mersin 10, Turkey.}
\pacs{PACS number}

\begin{abstract}
We consider the standard Yang-Mills (YM) invariant raised to the power q,
i.e., $(F_{\mu \nu }^{\left( a\right) }F^{\left( a\right) \mu \nu })^{q}$ as
the source of our geometry and investigate the possible black hole
solutions. How does this parameter q modify the black holes in
Einstein-Yang-Mills (EYM) and its extensions such as Gauss-Bonnet (GB) and
the third order Lovelock theories? The advantage of such a power q (or a set
of superposed members of the YM hierarchies) if any, may be tested even in a
free YM theory in flat spacetime. Our choice of the YM field is purely
magnetic in any higher dimensions so that duality makes no sense. In analogy
with the Einstein-power-Maxwell theory, the conformal invariance provides
further reduction, albeit in a spacetime for dimensions of multiples of 4.
\end{abstract}

\maketitle

\section{Introduction}

$N-$dimensional static, spherically symmetric Einstein-Yang-Mills (EYM)
black hole solutions in general relativity are well-known by now for which
we refer to \cite{1}, and references cited therein. YM theory's
non-linearity naturally adds further complexity to the already non-linear
gravity, thus expectedly the theory and its accompanied solutions become
rather complicated. Extension of the Einstein-Hilbert (EH) action with
further non-linearities, such as Gauss-Bonnet (GB) or Lovelock have also
been considered. These latter theories involve higher order invariants in
such combinations that the field equations remain second order.

More recently there has been aroused interest about black hole solutions
whose source is a power of the Maxwell scalar i.e., $(F_{\mu \nu }F^{\mu \nu
})^{q}$ , where $q$ is an arbitrary positive real number \cite{2}.
Subsequently this will be developed easily into a hierarchies of YM terms.
In the standard Maxwell theory we have $q=1$, whereas now the choice $q\neq 1
$ is also taken into account which adds to the theory a new dimension of
non-linearity from the electromagnetism. Non-linear electrodynamics, such as
Born-Infeld (BI) involves a kind of non-linearity that is more familiar for
a long time \cite{3}. From the outset we express that the non-linearity
involved in the power-Maxwell formalism is radically different from that of
BI. An infinite series expansion of the square root term in the latter
reveals this fact. For the special choice $q=\frac{N}{4},$ where $N=$%
dimension of the spacetime is a multiple of $4$, it yields a traceless
Maxwell's energy-momentum tensor which leads to conformal invariance. That
is, in the absence of different fields such as self-interacting massless
scalar field and /or a cosmological constant we have a vanishing scalar
curvature. This implies a relatively simpler geometry under the invariance $%
g_{\mu \nu }\rightarrow \Omega ^{2}g_{\mu \nu }$ and naturally attracts
interest. The absence of black hole solutions in higher dimensions for a
self-interacting scalar field was proved long time ago \cite{4}.
Self-interacting Maxwell field with a power of invariant, however, which
conformally interacts with gravity admits black hole solutions \cite{2}.

Being motivated by the black holes sourced by the power of Maxwell's
invariant we investigate in this work the existence of black holes with a
power of YM source. That is we shall choose our source as $\left( F_{\mu \nu
}^{\left( a\right) }F^{\left( a\right) \mu \nu }\right) ^{q}$, (and also $%
\overset{q}{\underset{k=0}{\tsum }b_{k}}\left( F_{\mu \nu }^{\left( a\right)
}F^{\left( a\right) \mu \nu }\right) ^{k},$ with constant coefficients $b_{k}
$) where $F_{\mu \nu }^{\left( a\right) }$ is the YM field with its internal
index $1\leq a\leq \frac{1}{2}\left( N-1\right) \left( N-2\right) $ and $q$
is a real number such that $q=1$ recovers the EYM black holes. Similar to
the power-Maxwell case we obtain the conformally invariant YM black holes
with a zero trace for the energy-momentum tensor. It turns out in analogy
that the dimensions of spacetimes are multiples of 4. The power $q$ can be
chosen arbitrary provided the conformal property is lost. It will also be
shown that $q<0$, will lead to the violation of the energy and causality
conditions. This will restrict us only to the choice $q>0$. As before, our
magnetically charged YM field consists of the Wu-Yang ansatz in any higher
dimensions\cite{1}. The EYM metric function admits an integral proportional
to $\sim \frac{\ln r}{r^{2}}$ for $N=5$ and $\sim \frac{1}{r^{2}}$ for all $%
N>5.$ The fixed $r-$dependence for $N>5,$ was considered to be unusual i.e.,
a drawback or advantage, depending on the region of interest. Now with the
choice of the power $q$ on the YM invariant we obtain dependence on $q$ as
well, which brings extra $r-$dependence in the metric. The possible set of
integer $q$ values in each $N>5$ is determined by the validity of the energy
conditions. For $N=4$ and $5$ we show that $q=1$, necessarily, but for $N>5$
we can't accommodate $q=1$ unless we violate some energy conditions.

We consider next the GB (i.e., second order Lovelock) and successively
Lovelock's third order term added to the first order EH Lagrangian. Our
source term throughout the paper is the YM invariant raised to the power $q$
(and its hierarchies). In each case, separately or together, we seek
solutions to what we call, the Einstein-power-YM (EPYM) field with GB and
Lovelock terms. It is remarkable that such a highly nonlinear theory with
non-linearities in various forms admits black hole solutions and in the
appropriate limits, with $q=1,$ it yields all the previously known
solutions. In the presence of both the second and third order Lovelock
terms, however, we impose for technical reasons an algebraic condition
between their parameters. This we do for the simple reason that the most
general solution involving both the second and third order terms is
technically far from being tractable. Useful thermodynamic quantities such
as the Hawking temperature, specific heat and free energy are determined and
briefly discussed.

Organization of the Letter is as follows. Sec. II contains the action, field
equations, energy-momentum for EPYM gravity and solutions to the field
equations. Sec.s III and IV follow a similar pattern for the GB and third
order Lovelock theories, respectively. Yang-Mills hierarchies are discussed
in Sec. V. We complete the Letter with Conclusion which appears in Sec. VI.

\section{Field equations and the metric ansatz for EPYM gravity}

The $N\left( =n+2\right) -$dimensional action for Einstein-power-Yang-Mills
(EPYM) gravity with a cosmological constant $\Lambda $ is given by ($8\pi
G=1 $) 
\begin{equation}
I=\frac{1}{2}\int_{\mathcal{M}}d^{n+2}x\sqrt{-g}\left( R-\frac{n\left(
n+1\right) }{3}\Lambda -\mathcal{F}^{q}\right) ,
\end{equation}%
in which $\mathcal{F}$ is the YM invariant%
\begin{eqnarray}
\mathcal{F} &=&\mathbf{Tr}(F_{\lambda \sigma }^{\left( a\right) }F^{\left(
a\right) \lambda \sigma }), \\
\mathbf{Tr}(.) &=&\overset{n(n+1)/2}{\underset{a=1}{\tsum }\left( .\right) },
\notag
\end{eqnarray}%
$R$ is the Ricci Scalar and $q$ is a positive real parameter. Here the YM
field is defined as 
\begin{equation}
\mathbf{F}^{\left( a\right) }=\mathbf{dA}^{\left( a\right) }+\frac{1}{%
2\sigma }C_{\left( b\right) \left( c\right) }^{\left( a\right) }\mathbf{A}%
^{\left( b\right) }\wedge \mathbf{A}^{\left( c\right) }
\end{equation}%
in which $C_{\left( b\right) \left( c\right) }^{\left( a\right) }$ stands
for the structure constants of $\frac{n(n+1)}{2}-$ parameter Lie group $G,$ $%
\sigma $ is a coupling constant and $\mathbf{A}^{\left( a\right) }$ are the $%
SO(n+1)$ gauge group YM potentials. The determination of the components $%
C_{\left( b\right) \left( c\right) }^{\left( a\right) }$ has been described
elsewhere\cite{5}. We note that the internal indices $\{a,b,c,...\}$ do not
differ whether in covariant or contravariant form. Variation of the action
with respect to the spacetime metric $g_{\mu \nu }$ yields the field
equations%
\begin{gather}
G_{\ \nu }^{\mu }+\frac{n\left( n+1\right) }{6}\Lambda \delta _{\ \nu }^{\mu
}=T_{\ \nu }^{\mu }, \\
T_{\ \nu }^{\mu }=-\frac{1}{2}\left( \delta _{\ \nu }^{\mu }\mathcal{F}%
^{q}-4q\mathbf{Tr}\left( F_{\nu \lambda }^{\left( a\right) }F^{\left(
a\right) \ \mu \lambda }\right) \mathcal{F}^{q-1}\right) ,
\end{gather}%
where $G_{\mu \nu }$ is the Einstein tensor. Variation with respect to the
gauge potentials $\mathbf{A}^{\left( a\right) }$ yields the YM equations%
\begin{equation}
\mathbf{d}\left( ^{\star }\mathbf{F}^{\left( a\right) }\mathcal{F}%
^{q-1}\right) +\frac{1}{\sigma }C_{\left( b\right) \left( c\right) }^{\left(
a\right) }\mathcal{F}^{q-1}\mathbf{A}^{\left( b\right) }\wedge ^{\star }%
\mathbf{F}^{\left( c\right) }=0,
\end{equation}%
where $^{\star }$ means duality. It is readily observed that for $q=1$ our
formalism reduces to the standard EYM theory. Our objective in this work
therefore is to study the role of the parameter $q$ in the black holes. Our
metric ansatz for $N\left( =n+2\right) $ dimensions, is chosen as 
\begin{equation}
ds^{2}=-f\left( r\right) dt^{2}+\frac{dr^{2}}{f\left( r\right) }%
+r^{2}d\Omega _{n}^{2},
\end{equation}%
in which $f\left( r\right) $ is our metric function and 
\begin{equation}
d\Omega _{n}^{2}=d\theta _{1}^{2}+\underset{i=2}{\overset{n}{\tsum }}%
\underset{j=1}{\overset{i-1}{\tprod }}\sin ^{2}\theta _{j}\;d\theta _{i}^{2},
\end{equation}%
where%
\begin{equation*}
0\leq \theta _{n}\leq 2\pi ,0\leq \theta _{i}\leq \pi ,\text{ \ \ }1\leq
i\leq n-1.
\end{equation*}%
The choice of these metrics can be traced back to the form of the
stress-energy tensor (5), which satisfies $T_{0}^{0}-T_{1}^{1}=0$ (see Eq.
(12) below) and consequently $G_{0}^{0}-G_{1}^{1}=0$, whose explicit form,
on integration, gives $\left\vert g_{00}g_{11}\right\vert =C=$constant. We
need only to choose the time scale at infinity to make this constant equal
to unity.

\subsection{Energy momentum tensor}

Recently we have introduced and used the higher dimensional version of the
Wu-Yang ansatz in EYM theory of gravity \cite{1,5}. In this ansatz we
express the Yang-Mills magnetic gauge potential one-forms as%
\begin{align}
\mathbf{A}^{(a)}& =\frac{Q}{r^{2}}C_{\left( i\right) \left( j\right)
}^{\left( a\right) }\ x^{i}dx^{j},\text{ \ \ }Q=\text{YM magnetic charge, \ }%
r^{2}=\overset{n+1}{\underset{i=1}{\sum }}x_{i}^{2}, \\
2& \leq j+1\leq i\leq n+1,\text{ \ and \ }1\leq a\leq n(n+1)/2,  \notag \\
x_{1}& =r\cos \theta _{n-1}\sin \theta _{n-2}...\sin \theta _{1},\text{ }%
x_{2}=r\sin \theta _{n-1}\sin \theta _{n-2}...\sin \theta _{1},  \notag \\
\text{ }x_{3}& =r\cos \theta _{n-2}\sin \theta _{n-3}...\sin \theta _{1},%
\text{ }x_{4}=r\sin \theta _{n-2}\sin \theta _{n-3}...\sin \theta _{1}, 
\notag \\
& ...  \notag \\
x_{n}& =r\cos \theta _{1}.  \notag
\end{align}%
One can easily show that these ansaetze satisfy the YM equations\cite{1,5}.
In consequence, the energy momentum tensor (5), with 
\begin{eqnarray}
\mathcal{F} &=&\frac{n\left( n-1\right) Q^{2}}{r^{4}}, \\
\mathbf{Tr}\left( F_{\theta _{i}\lambda }^{\left( a\right) }F^{\left(
a\right) \ \theta _{i}\lambda }\right)  &=&\frac{\left( n-1\right) Q^{2}}{%
r^{4}}=\frac{1}{n}\mathcal{F}
\end{eqnarray}%
becomes%
\begin{equation}
T_{\text{ }b}^{a}=-\frac{1}{2}\mathcal{F}^{q}\text{diag}\left[ 1,1,\kappa
,\kappa ,..,\kappa \right] ,\text{ \ and \ }\kappa =\left( 1-\frac{4q}{n}%
\right) .
\end{equation}%
We observe that the trace of $T_{\text{ }b}^{a}$ is $T=-\frac{1}{2}\mathcal{F%
}^{q}\left( N-4q\right) $ which vanishes for the particular case $q=\frac{N}{%
4}.$ It is also remarkable to give the intervals of $q$ in which the Weak
Energy Condition (WEC), Strong Energy Condition (SEC), Dominant Energy
Condition (DEC) and Causality Condition (CC) are satisfied \cite{6}. It is
observed from Tab. (1) that the physically meaningful range for $q$ is $%
\frac{n+1}{4}\leq q<\frac{n+1}{2},$ which satisfies all the energy and
causality conditions. The choice $q<0,$ violates all these conditions so it
must be discarded. In the sequel we shall use this energy momentum tensor to
find black hole solutions for the EPYM, EPYMGB and EPYMGBL field equations
with the cosmological constant $\Lambda .$

\subsection{EPYM Black hole solution for $N\geq 5$ dimensions}

In $N\left( =n+2\right) \geq 5$ dimensions the $rr$ component of Einstein
equation reads%
\begin{equation}
\frac{3\left( n\left( n-1\right) Q^{2}\right) ^{q}}{r^{2\left( 2q-1\right) }}%
+3n\left[ rg^{\prime }\left( r\right) +\left( n-1\right) g\left( r\right) +%
\frac{\Lambda }{3}\left( n+1\right) r^{2}\right] =0,
\end{equation}%
in which $f\left( r\right) =1+g\left( r\right) .$ Direct integration leads
to the following solutions 
\begin{eqnarray}
f\left( r\right) &=&\left\{ 
\begin{tabular}{ll}
$1-\frac{4m}{nr^{n-1}}-\frac{Q_{1}}{r^{4q-2}}-\frac{\Lambda }{3}r^{2},$ & $%
q\neq \frac{n+1}{4}$ \\ 
$1-\frac{4m}{nr^{n-1}}-\frac{Q_{2}\ln r}{r^{n-1}}-\frac{\Lambda }{3}r^{2},$
& $q=\frac{n+1}{4}$%
\end{tabular}%
\right. \\
Q_{1} &=&\frac{\left( \left( n-1\right) nQ^{2}\right) ^{q}}{n\left(
n+1-4q\right) },Q_{2}=\frac{\left( \left( n-1\right) nQ^{2}\right) ^{\frac{%
n+1}{4}}}{n}  \notag
\end{eqnarray}%
where $m$ is the ADM mass of the black hole. It is observed that physical
properties of such a black hole depends on the parameter $q$. The location
of horizons, $f\left( r_{h}\right) =0,$ involves an algebraic equation whose
roots can be found numerically. The entropy, Hawking temperature and other
thermodynamics properties all can be calculated accordingly and they are
dependent on $q$. Tab. (1) shows that the minimum possible value for $q$
which provides all the energy conditions to be satisfied is given by $%
q_{\min }=\frac{n+1}{4},$ that is, the case of solution with logaritmic
term. In $5-$dimensions $q_{\min }=1,$ which recovers the usual EYM solution
found in\cite{1,5}. With the exception of $N=5$ where $q=1$ is part of
possible $q^{\prime }s$ (which satisfy all the energy conditions), in higher
dimensions $q$ must be greater than one. For instance, in $6-$dimensions $%
\frac{5}{4}\leq q<\frac{5}{2}$ and in $7-$dimensions $\frac{3}{2}\leq q<3.$
If one constrains $q$ to be an integer, Tab. (2) gives the possible $q$
values in some dimensions. From this table we can identify the dimensions in
which the logarithmic term appears naturally. These are $N=5,9,13,...,$ for
which $q_{\min }=\frac{N-1}{4}$ is an integer. Let us remark that since for $%
N=4$ our YM field gauge transforms to an Abelian form\cite{7}, our results
become automatically valid also for $N=4$.

We observe that although the metric function $f\left( r\right) $ at infinity
goes to $-\frac{\Lambda }{3}r^{2}$ its behavior about the origin is quite
different and strongly depends on $q$ i.e.,%
\begin{equation}
\lim_{r\rightarrow 0}f\left( r\right) \rightarrow \left\{ 
\begin{tabular}{ll}
$-\frac{4m}{nr^{n-1}}\rightarrow -\infty ,$ & $q<\frac{n+1}{4}$ \\ 
$\frac{\left( \left( n-1\right) nQ^{2}\right) ^{\frac{n+1}{4}}}{nr^{n-1}}\ln
\left( \frac{1}{r}\right) \rightarrow +\infty ,$ & $q=\frac{n+1}{4}$ \\ 
$\frac{\left( \left( n-1\right) nQ^{2}\right) ^{q}}{n\left( 4q-n-1\right)
r^{4q-2}}\rightarrow +\infty ,$ & $q>\frac{n+1}{4}$%
\end{tabular}%
\right. .
\end{equation}%
This is important because for the case of $q\geq \frac{n+1}{4}$ one may
adjust the mass and charge to have a metric function in contradiction with
the Cosmic Censorship Conjecture (CCC). One statement of this conjecture is
that all singularities (here $r=0$) are hidden behind event horizons. Of
course, nature may restrict $Q$ and $m$ in order not to violate this
conjecture.

Note that $r=0$ is a singularity for the metric whose Ricci scalar is given
by%
\begin{equation}
R=\left\{ 
\begin{tabular}{ll}
$\frac{\left( \left( n-1\right) nQ^{2}\right) ^{q}\left( n+2-4q\right) }{%
nr^{4q}},$ & $q\neq \frac{n+1}{4}$ \\ 
$\frac{\left( \left( n-1\right) nQ^{2}\right) ^{\frac{n+1}{4}}}{nr^{n+1}},$
& $q=\frac{n+1}{4}$%
\end{tabular}%
\right. .
\end{equation}

\subsection{Extremal Black Holes}

Closely related with a usual black hole is an extremal black hole whose
horizons coincide. As it is well known to get extremal solution one should
solve $f\left( r\right) =0,$ and $f^{\prime }\left( r\right) =0$
simultaneously. This set of equations for the solution (14), without
cosmological constant, leads to 
\begin{eqnarray}
r_{e} &=&\left( n\left( n-1\right) \right) ^{\frac{q-1}{2\left( 2q-1\right) }%
}Q^{\left( \frac{q}{2q-1}\right) }, \\
m_{e} &=&\left\{ 
\begin{tabular}{ll}
$\frac{\left( 2q-1\right) Q^{\left( \frac{q}{2q-1}\right) }}{2\left(
4q-n-1\right) }n^{\frac{3q-1-n+qn}{2\left( 2q-1\right) }}\left( n-1\right) ^{%
\frac{\left( n-1\right) \left( q-1\right) }{2\left( 2q-1\right) }},$ & $%
q\neq \frac{n+1}{4}$ \\ 
$\frac{\left( n-1\right) ^{\frac{n-3}{4}}Q^{\frac{n+1}{2}}}{8}\left\{ n^{%
\frac{n+1}{4}}\left( 2+\frac{3}{2}\ln \frac{n\left( n-1\right) }{Q^{2/3}}%
\right) -\frac{n^{\frac{n+5}{4}}}{2}\ln \left( n\left( n-1\right)
Q^{2}\right) \right\} ,$ & $q=\frac{n+1}{4}$%
\end{tabular}%
\right.
\end{eqnarray}%
where $r_{e}$ is the radius of degenerate horizon and $m_{e}$ and $Q$ are
the extremal mass and charge of the black hole, respectively. One may check
the case of $q=1,$ resulting in%
\begin{equation}
r_{e}=Q,\text{ \ \ }m_{e}=\frac{n}{2\left( 3-n\right) }Q,
\end{equation}%
which clearly in 4-dimensions gives $r_{e}=Q=$\ \ $m_{e},$ as it should.

\subsection{Thermodynamics of the EPYM black hole}

In this section we present some thermodynamical properties of EPYM black
hole solution with cosmological constant. Here it is convenient to rescale
our quantities in terms of some different powers of radius of the horizon $%
r_{h},$ i.e., we introduce 
\begin{equation}
\check{T}_{H}=T_{H}r_{h},\text{ }\check{M}_{ADM}=M_{ADM}/r_{h}^{n-1},\text{ }%
\check{\Lambda}=\Lambda r_{h}^{2},\text{ }\check{Q}_{i}=Q_{i}/r_{h}^{2\left(
2q-1\right) },\text{ }\check{C}=C/r_{h}^{n}\text{ and }\check{F}=F/r^{n-1},
\end{equation}%
where $T_{H}=f^{\prime }\left( r_{h}\right) /4\pi $ is the Hawking
temperature, $C=C_{Q}=T_{H}\left( \frac{\partial S}{\partial T_{H}}\right)
_{Q}$ is the heat capacity for constant $Q$ and $F=M_{ADM}-T_{H}S$ is the
free energy of the black hole as a thermodynamical system. Therein%
\begin{equation}
S=\frac{A}{4}=\frac{\left( n+1\right) \pi ^{\left( \frac{n+1}{2}\right) }}{%
4\Gamma \left( \frac{n+3}{2}\right) }r_{h}^{n}
\end{equation}%
is the Bekenstein-Hawking entropy where $\Gamma \left( .\right) $ stands for
the gamma function. As one may notice in (14) $m$ represent the ADM mass of
the black hole. This helps us to write 
\begin{equation}
\check{M}_{ADM}=\check{m}=\left\{ 
\begin{tabular}{ll}
$\frac{n}{4}\left( 1-\check{Q}_{1}-\frac{\check{\Lambda}}{3}\right) ,$ & $%
q\neq \frac{n+1}{4}$ \\ 
$\frac{n}{4}\left( 1-\check{Q}_{2}\ln \left( r_{h}\right) -\frac{\check{%
\Lambda}}{3}\right) ,$ & $q=\frac{n+1}{4}$%
\end{tabular}%
\right.
\end{equation}%
which imposes some restrictions on $\check{Q}_{i}$\ and $\check{\Lambda}$ in
order to have a positive and physically acceptable $\check{M}_{ADM}.$

In terms of the event horizon $r_{h}$ Hawking temperature becomes 
\begin{equation}
\check{T}_{H}=\left\{ 
\begin{tabular}{ll}
$\frac{-\check{Q}_{1}\left( n+1-4q\right) -\frac{\check{\Lambda}}{3}\left(
n+1\right) +\left( n-1\right) }{4\pi },$ & $q\neq \frac{n+1}{4}$ \\ 
$\frac{-\check{Q}_{2}-\frac{\check{\Lambda}}{3}\left( n+1\right) +\left(
n-1\right) }{4\pi },$ & $q=\frac{n+1}{4}$%
\end{tabular}%
\right. .
\end{equation}%
For the case of $q\neq \frac{n+1}{4}$ clearly by imposing $\check{M}_{ADM},%
\check{T}_{H}>0$ one finds $\frac{\check{\Lambda}}{3}<\left( 1-\frac{1}{2q}%
\right) $ and for the case of $q=\frac{n+1}{4}$ and choosing $r_{h}=1,$ one
gets $\frac{\check{\Lambda}}{3}<1-\frac{\check{Q}_{2}+2}{n+1}.$ The heat
capacity $\check{C}$ is given by

\begin{equation}
\check{C}=\left\{ 
\begin{tabular}{ll}
$\frac{n}{2}\frac{\pi ^{\frac{n+1}{2}}}{\Gamma \left( \frac{n+1}{2}\right) }%
\frac{\frac{\left( n+1\right) \check{\Lambda}}{3}+\left( n+1-4q\right) 
\check{Q}_{1}-\left( n-1\right) }{\frac{\check{\Lambda}}{3}\left( n+1\right)
-\check{Q}_{1}\left( n+1-4q\right) \left( 4q-1\right) +n-1},$ & $q\neq \frac{%
n+1}{4}$ \\ 
$\frac{n}{2}\frac{\pi ^{\frac{n+1}{2}}}{\Gamma \left( \frac{n+1}{2}\right) }%
\frac{\frac{\left( n+1\right) \check{\Lambda}}{3}-\left( n-1\right) +\check{Q%
}_{2}}{\left( n+1\right) \frac{\check{\Lambda}}{3}-n\check{Q}_{2}+\left(
n-1\right) },$ & $q=\frac{n+1}{4}$%
\end{tabular}%
\right.
\end{equation}%
which reveals the thermodynamic instability of the black hole. In fact the
possible roots of denominator of $\check{C}$ present a phase transition
which can be interpreted as thermodynamical instability.

For completeness we give also the free energy $F$ of our black hole as a
thermodynamical system, which is%
\begin{equation}
\check{F}=\left\{ 
\begin{tabular}{ll}
$\frac{\left[ \check{Q}_{1}\left( n+1-4q\right) +\frac{\check{\Lambda}\left(
n+1\right) }{3}+1-n\right] \pi ^{\frac{n-1}{2}}-2n\left( \check{Q}_{1}+\frac{%
\check{\Lambda}}{3}-1\right) \Gamma \left( \frac{n+1}{2}\right) }{8\Gamma
\left( \frac{n+1}{2}\right) },$ & $q\neq \frac{n+1}{4}$ \\ 
$\frac{\left[ \check{Q}_{2}+\frac{\check{\Lambda}\left( n+1\right) }{3}%
-\left( n-1\right) \right] \pi ^{\frac{n-1}{2}}-2n\left( \check{Q}_{2}\ln
\left( r_{h}\right) +\left( \frac{\check{\Lambda}}{3}-1\right) \right)
\Gamma \left( \frac{n+1}{2}\right) }{8\Gamma \left( \frac{n+1}{2}\right) },$
& $q=\frac{n+1}{4}$%
\end{tabular}%
\right. .
\end{equation}%
By letting $q=1$ and $n=2$ for the $4-$dimensional Reissner-Nordstr\"{o}m
metric, the foregoing expressions become%
\begin{eqnarray}
M_{ADM} &=&m=\frac{r_{h}}{2}\left( 1+\frac{Q^{2}}{r_{h}^{2}}\right) , \\
T_{H} &=&\frac{f^{\prime }\left( r_{h}\right) }{4\pi }=\frac{1}{4\pi r_{h}}%
\left( 1-\frac{Q^{2}}{r_{h}^{2}}\right) , \\
S_{BH} &=&\pi r_{h}^{2}, \\
C_{Q} &=&-\frac{2\pi \left[ 1-\frac{Q^{2}}{r_{h}^{2}}\right] r_{h}^{2}}{%
\left[ 1-3\frac{Q^{2}}{r_{h}^{2}}\right] }, \\
F &=&\left( 1+\frac{3Q^{2}}{r_{h}^{2}}\right) \frac{r_{h}}{4}.
\end{eqnarray}

\section{Field Equations and the metric ansatz for EPYMGB gravity}

The EPYMGB action in $N(=n+2)-$dimensions is given by ($8\pi G=1$)%
\begin{equation}
I=\frac{1}{2}\int_{\mathcal{M}}d^{n+2}x\sqrt{-g}\left( R-\frac{n\left(
n+1\right) }{3}\Lambda +\alpha \mathcal{L}_{GB}-\mathcal{F}^{q}\right) ,
\end{equation}%
where $\alpha $ is the GB parameter and $\mathcal{L}_{GB}$ is given by 
\begin{equation}
\mathcal{L}_{GB}=R_{\mu \nu \gamma \delta }R^{\mu \nu \gamma \delta
}-4R_{\mu \nu }R^{\mu \nu }+R^{2}.
\end{equation}%
Variation of the new action with respect to the space-time metric $g_{\mu
\nu }$ yields the field equations%
\begin{equation}
G_{\mu \nu }^{E}+\alpha G_{\mu \nu }^{GB}+\frac{n\left( n+1\right) }{6}%
\Lambda g_{\mu \nu }=T_{\mu \nu },
\end{equation}%
where 
\begin{equation}
G_{\mu \nu }^{GB}=2\left( -R_{\mu \sigma \kappa \tau }R_{\quad \nu }^{\kappa
\tau \sigma }-2R_{\mu \rho \nu \sigma }R^{\rho \sigma }-2R_{\mu \sigma }R_{\
\nu }^{\sigma }+RR_{\mu \nu }\right) -\frac{1}{2}\mathcal{L}_{GB}g_{\mu \nu }%
\text{ ,}
\end{equation}%
and $T_{\mu \nu }$ is given by (12).

\subsection{EPYMGB Black hole solution for $N\geq 5$ dimensions}

As before, the $rr$ component of Einstein equation (33) can be written as%
\begin{gather}
\frac{3\left( n\left( n-1\right) Q^{2}\right) ^{q}}{r^{4\left( q-1\right) }}%
+3n\left[ \left( r^{3}-2\tilde{\alpha}_{2}rg\left( r\right) \right)
g^{\prime }\left( r\right) -\tilde{\alpha}_{2}\left( n-2\right) g\left(
r\right) ^{2}\right. +  \notag \\
\left. r^{2}\left( n-1\right) g\left( r\right) +\frac{\Lambda }{3}\left(
n+1\right) r^{4}\right] =0,
\end{gather}%
in which $\widetilde{\alpha }_{2}=\left( n-1\right) \left( n-2\right) \alpha
_{2}.$ This equation admits a solution as%
\begin{equation}
f_{\pm }\left( r\right) =\left\{ 
\begin{tabular}{ll}
$1+\frac{r^{2}}{2\tilde{\alpha}_{2}}\left( 1\pm \sqrt{1+\frac{4}{3}\Lambda 
\tilde{\alpha}_{2}+\frac{16m\tilde{\alpha}_{2}}{nr^{n+1}}+\frac{4\tilde{%
\alpha}_{2}Q_{1}}{r^{4q}}}\right) ,$ & $q\neq \frac{n+1}{4}$ \\ 
$1+\frac{r^{2}}{2\tilde{\alpha}_{2}}\left( 1\pm \sqrt{1+\frac{4}{3}\Lambda 
\tilde{\alpha}_{2}+\frac{16m\tilde{\alpha}_{2}}{nr^{n+1}}+\frac{4\tilde{%
\alpha}_{2}Q_{2}\ln \left( r\right) }{r^{n+1}}}\right) ,$ & $q=\frac{n+1}{4}$%
\end{tabular}%
\right. .
\end{equation}%
The asymptotic behavior of the metric reveals that%
\begin{equation}
\lim_{r\rightarrow \infty }f_{\pm }\left( r\right) \rightarrow \left\{ 
\begin{tabular}{ll}
$1+\frac{r^{2}}{2\tilde{\alpha}_{2}}\left( 1\pm \sqrt{1+\frac{4}{3}\Lambda 
\tilde{\alpha}_{2}}\right) ,$ & $q<\frac{n+1}{4}$ \\ 
$1+\frac{r^{2}}{2\tilde{\alpha}_{2}}\left( 1\pm \sqrt{1+\frac{4}{3}\Lambda 
\tilde{\alpha}_{2}}\right) ,$ & $q=\frac{n+1}{4}$ \\ 
$1+\frac{r^{2}}{2\tilde{\alpha}_{2}}\left( 1\mp \sqrt{1+\frac{4}{3}\Lambda 
\tilde{\alpha}_{2}}\right) ,$ & $q>\frac{n+1}{4}$%
\end{tabular}%
\right. ,
\end{equation}%
which depending on $\Lambda $ it is de Sitter, Anti de Sitter or flat.
Abiding by the (anti) de Sitter limit for $\tilde{\alpha}_{2}\rightarrow 0,$
we must choose the $\left( -\right) $ sign.

\subsection{Thermodynamics of the EPYMGB black hole}

By using the above rescaling plus $\check{\alpha}_{2}=\tilde{\alpha}%
_{2}/r_{h}^{2},$ one can find the Hawking temperature of the EPYMGB black
hole solutions (36) as 
\begin{eqnarray}
\check{T}_{H}\left( -\right) &=&\left\{ 
\begin{tabular}{ll}
$-\frac{\check{Q}_{1}\left( n+1-4q\right) +\frac{\check{\Lambda}}{3}\left(
n+1\right) -\left( n-1\right) -\check{\alpha}_{2}\left( n-3\right) }{4\pi
\left( 1+2\check{\alpha}_{2}\right) }$ & $q\neq \frac{n+1}{4}$ \\ 
$-\frac{\check{Q}_{2}+\frac{\check{\Lambda}}{3}\left( n+1\right) -\left(
n-1\right) -\check{\alpha}_{2}\left( n-3\right) }{4\pi \left( 1+2\check{%
\alpha}_{2}\right) }$ & $q=\frac{n+1}{4}$%
\end{tabular}%
\right. , \\
\check{T}_{H}\left( +\right) &=&\left\{ 
\begin{tabular}{ll}
$\frac{\check{Q}_{1}\check{\alpha}_{2}\left( n+1-4q\right) +\frac{\check{%
\Lambda}}{3}\check{\alpha}_{2}\left( n+1\right) -\check{\alpha}%
_{2}^{2}\left( n-3\right) -\check{\alpha}_{2}\left( n-5\right) +2}{4\pi 
\check{\alpha}_{2}\left( 1+2\check{\alpha}_{2}\right) }$ & $q\neq \frac{n+1}{%
4}$ \\ 
$\frac{\check{Q}_{2}\check{\alpha}_{2}+\frac{\check{\Lambda}}{3}\check{\alpha%
}_{2}\left( n+1\right) -\check{\alpha}_{2}^{2}\left( n-3\right) -\check{%
\alpha}_{2}\left( n-5\right) +2}{4\pi \check{\alpha}_{2}\left( 1+2\check{%
\alpha}_{2}\right) }$ & $q=\frac{n+1}{4}$%
\end{tabular}%
\right. ,
\end{eqnarray}%
here $\left( \pm \right) $ state the correspondence branches. Here we
observe that $\check{T}_{H}\left( -\right) $ in the limit of $\check{\alpha}%
_{2}\rightarrow 0$ correctly reduces to the Hawking temperature of EPYM
black hole (23) as expected. It is remarkable to observe that $\check{\alpha}%
_{2}=-\frac{1}{2}$ is a point of infinite temperature, or instability of the
black hole. This means that if $\tilde{\alpha}_{2}/r_{h}^{2}=-\frac{1}{2},$
the black hole will be unstable. For the positive branch one should be
careful about $\check{\alpha}_{2}\rightarrow 0$ which is not applicable.

In the sequel we give the other thermodynamical properties of the BH
solution (36) in separate cases.

\subsubsection{Negative branch $q\neq \frac{n+1}{4}$}

The ADM mass:%
\begin{equation}
\check{M}_{ADM}=\check{m}=\frac{n}{4}\left( 1+\check{\alpha}_{2}-\check{Q}%
_{1}-\frac{\check{\Lambda}}{3}\right) ,
\end{equation}%
The heat capacity: 
\begin{equation}
\begin{tabular}{l}
$\check{C}=\frac{n}{2}\pi ^{\frac{n+1}{2}}\frac{\left( \check{\alpha}_{2}+%
\frac{1}{2}\right) }{\Gamma \left( \frac{n+1}{2}\right) }\frac{\frac{\check{%
\Lambda}}{3}\left( n+1\right) +\left( n+1-4q\right) \check{Q}_{1}-\left(
n-3\right) \check{\alpha}_{2}-\left( n-1\right) }{\left\{ \left( n+1\right)
\left( \check{\alpha}_{2}+\frac{1}{6}\right) \check{\Lambda}+\left(
n-3\right) \check{\alpha}_{2}^{2}-4\left[ \left( q-\frac{3}{4}\right) \check{%
\alpha}_{2}+\frac{1}{2}\left( q-\frac{1}{4}\right) \right] \check{Q}%
_{1}\left( 1+n-4q\right) +\left( \frac{n-7}{2}\right) \check{\alpha}%
_{2}+\left( \frac{n-1}{2}\right) \right\} },$%
\end{tabular}%
\end{equation}%
The free energy:%
\begin{equation}
\begin{tabular}{l}
$\check{F}=\frac{\left[ \check{Q}_{1}\left( n+1-4q\right) +\frac{\check{%
\Lambda}\left( n+1\right) }{3}-\left( n-3\right) \check{\alpha}_{2}+1-n%
\right] \pi ^{\frac{n-1}{2}}-4n\left( \check{Q}_{1}+\frac{\check{\Lambda}}{3}%
-1-\check{\alpha}_{2}\right) \left( \check{\alpha}_{2}+\frac{1}{2}\right)
\Gamma \left( \frac{n+1}{2}\right) }{8\left( 1+2\check{\alpha}_{2}\right)
\Gamma \left( \frac{n+1}{2}\right) }.$%
\end{tabular}%
\end{equation}

\subsubsection{Negative branch $q=\frac{n+1}{4}$}

The ADM mass:%
\begin{equation}
\check{M}_{ADM}=\check{m}=\frac{n}{4}\left( 1+\check{\alpha}_{2}-\check{Q}%
_{2}\ln \left( r_{h}\right) -\frac{\check{\Lambda}}{3}\right) ,
\end{equation}%
The heat capacity: 
\begin{equation}
\begin{tabular}{l}
$\check{C}=n\pi ^{\frac{n+1}{2}}\frac{\left( \check{\alpha}_{2}+\frac{1}{2}%
\right) }{\Gamma \left( \frac{n+1}{2}\right) }\frac{\frac{\left( n+1\right) 
\check{\Lambda}}{3}-\left( n-1\right) -\left( n-3\right) \check{\alpha}_{2}+%
\check{Q}_{2}}{\left( n+1\right) \left( 1+6\check{\alpha}_{2}\right) \frac{%
\check{\Lambda}}{3}-\left( 2\left( n-2\right) \check{\alpha}_{2}+n\right) 
\check{Q}_{2}+2\left( n-3\right) \check{\alpha}_{2}^{2}+\left( n-7\right) 
\check{\alpha}_{2}+\left( n-1\right) },$%
\end{tabular}%
\end{equation}%
The free energy:%
\begin{equation}
\begin{tabular}{l}
$\check{F}=\frac{\left[ \check{Q}_{2}+\frac{\check{\Lambda}\left( n+1\right) 
}{3}-\left( n-3\right) \check{\alpha}_{2}-\left( n-1\right) \right] \pi ^{%
\frac{n-1}{2}}-4n\left( \check{Q}_{2}\ln \left( r_{h}\right) +\left( \frac{%
\check{\Lambda}}{3}-1\right) -\check{\alpha}_{2}\right) \left( \check{\alpha}%
_{2}+\frac{1}{2}\right) \Gamma \left( \frac{n+1}{2}\right) }{8\left( 1+2%
\check{\alpha}_{2}\right) \Gamma \left( \frac{n+1}{2}\right) }.$%
\end{tabular}%
\end{equation}

\subsubsection{Positive branch $q\neq \frac{n+1}{4}$}

The ADM mass:%
\begin{equation}
\check{M}_{ADM}=\check{m}=\frac{n}{4}\left( 1+\check{\alpha}_{2}-\check{Q}%
_{1}-\frac{\check{\Lambda}}{3}\right) ,
\end{equation}%
The heat capacity: 
\begin{equation}
\begin{tabular}{l}
$\check{C}=-\frac{n}{2}\pi ^{\frac{n+1}{2}}\frac{\left( \check{\alpha}_{2}+%
\frac{1}{2}\right) }{\Gamma \left( \frac{n+1}{2}\right) }\frac{-\frac{\check{%
\Lambda}}{3}\left( n+1\right) \check{\alpha}_{2}-\left( n+1-4q\right) \check{%
\alpha}_{2}\check{Q}_{1}+\left( n-3\right) \check{\alpha}_{2}^{2}+\left(
n-5\right) \check{\alpha}_{2}-2}{\left\{ \left( n+1\right) \left( \check{%
\alpha}_{2}+\frac{1}{6}\right) \check{\alpha}_{2}\check{\Lambda}+\left(
n-3\right) \check{\alpha}_{2}^{3}-4\left[ \left( q-\frac{3}{4}\right) \check{%
\alpha}_{2}+\frac{1}{2}\left( q-\frac{1}{4}\right) \right] \check{\alpha}_{2}%
\check{Q}_{1}\left( 1+n-4q\right) +\left( \frac{n+1}{2}\right) \check{\alpha}%
_{2}^{2}+\left( \frac{n+7}{2}\right) \check{\alpha}_{2}+1\right\} },$%
\end{tabular}%
\end{equation}%
The free energy:%
\begin{equation}
\begin{tabular}{l}
$\check{F}=\frac{\left[ -\check{Q}_{1}\check{\alpha}_{2}\left( n+1-4q\right)
-\frac{\check{\Lambda}\left( n+1\right) }{3}\check{\alpha}_{2}+\left(
n-3\right) \check{\alpha}_{2}^{2}+\left( n-5\right) \check{\alpha}_{2}-2%
\right] \pi ^{\frac{n-1}{2}}-4n\check{\alpha}_{2}\left( \check{Q}_{1}+\frac{%
\check{\Lambda}}{3}-1-\check{\alpha}_{2}\right) \left( \check{\alpha}_{2}+%
\frac{1}{2}\right) \Gamma \left( \frac{n+1}{2}\right) }{8\check{\alpha}%
_{2}\left( 1+2\check{\alpha}_{2}\right) \Gamma \left( \frac{n+1}{2}\right) }%
. $%
\end{tabular}%
\end{equation}

\subsubsection{Positive branch $q=\frac{n+1}{4}$}

The ADM mass:%
\begin{equation}
\check{M}_{ADM}=\check{m}=\frac{n}{4}\left( 1+\check{\alpha}_{2}-\check{Q}%
_{2}\ln \left( r_{h}\right) -\frac{\check{\Lambda}}{3}\right) ,
\end{equation}%
The heat capacity: 
\begin{equation}
\begin{tabular}{l}
$\check{C}=n\pi ^{\frac{n+1}{2}}\frac{\left( \check{\alpha}_{2}+\frac{1}{2}%
\right) }{\Gamma \left( \frac{n+1}{2}\right) }\frac{\frac{\check{\Lambda}}{3}%
\left( n+1\right) \check{\alpha}_{2}-\left( n-5\right) \check{\alpha}%
_{2}-\left( n-3\right) \check{\alpha}_{2}^{2}+\check{Q}_{2}\check{\alpha}%
_{2}+2}{\left\{ 2\left( n+1\right) \left( \check{\alpha}_{2}+\frac{1}{6}%
\right) \check{\alpha}_{2}\check{\Lambda}+2\left( n-3\right) \check{\alpha}%
_{2}^{3}-\left[ 2\left( n-2\right) \check{\alpha}_{2}+n\right] \check{\alpha}%
_{2}\check{Q}_{2}+\left( n+1\right) \check{\alpha}_{2}^{2}+\left( n+7\right) 
\check{\alpha}_{2}+2\right\} },$%
\end{tabular}%
\end{equation}%
The free energy:%
\begin{equation}
\begin{tabular}{l}
$\check{F}=\frac{\left[ -\check{Q}_{2}\check{\alpha}_{2}-\frac{\check{\Lambda%
}\left( n+1\right) }{3}\check{\alpha}_{2}+\left( n-3\right) \check{\alpha}%
_{2}^{2}+\left( n-5\right) \check{\alpha}_{2}-2\right] \pi ^{\frac{n-1}{2}%
}-4n\check{\alpha}_{2}\left( \check{Q}_{2}\ln r_{h}+\left( \frac{\check{%
\Lambda}}{3}-1-\check{\alpha}_{2}\right) \right) \left( \check{\alpha}_{2}+%
\frac{1}{2}\right) \Gamma \left( \frac{n+1}{2}\right) }{8\check{\alpha}%
_{2}\left( 1+2\check{\alpha}_{2}\right) \Gamma \left( \frac{n+1}{2}\right) }%
. $%
\end{tabular}%
\end{equation}

Finally in this section we look at $\check{C}$ which clearly, in general,
vanishes at $\check{\alpha}_{2}=-\frac{1}{2}$. Also any possible root for
the denominator of $\check{C}$ gives instability point or a phase transition.

\section{Field Equations and the metric ansatz for EPYMGBL gravity}

In this section we consider a more general action which involves, beside the
GB term, the third order Lovelock term\cite{8,9}. The EPYMGBL action in $%
N(=n+2)-$dimensions is given by ($8\pi G=1$)%
\begin{equation}
I=\frac{1}{2}\int_{\mathcal{M}}d^{n+2}x\sqrt{-g}\left( R-\frac{n\left(
n+1\right) }{3}\Lambda +\alpha _{2}\mathcal{L}_{GB}+\alpha _{3}\mathcal{L}%
_{(3)}-\mathcal{F}^{q}\right) ,
\end{equation}%
where $\alpha _{2}$ and $\alpha _{3}$ are the second and third order
Lovelock parameters respectively, and\cite{8} 
\begin{align}
\mathcal{L}_{\left( 3\right) }& =2R^{\mu \nu \sigma \kappa }R_{\sigma \kappa
\rho \tau }R_{\quad \mu \nu }^{\rho \tau }+8R_{\quad \sigma \rho }^{\mu \nu
}R_{\quad \nu \tau }^{\sigma \kappa }R_{\quad \mu \kappa }^{\rho \tau } 
\notag \\
& +24R^{\mu \nu \sigma \kappa }R_{\sigma \kappa \nu \rho }R_{\ \mu }^{\rho
}+3RR^{\mu \nu \sigma \kappa }R_{\sigma \kappa \mu \nu }  \notag \\
& +24R^{\mu \nu \sigma \kappa }R_{\sigma \mu }R_{\kappa \nu }+16R^{\mu \nu
}R_{\nu \sigma }R_{\ \mu }^{\sigma } \\
& -12RR^{\mu \nu }R_{\mu \nu }+R^{3},  \notag
\end{align}%
is the third order Lovelock Lagrangian. Variation of the new action with
respect to the space-time metric $g_{\mu \nu }$ yields the field equations%
\begin{equation}
G_{\mu \nu }+\alpha _{2}G_{\mu \nu }^{GB}+\alpha _{3}G_{\mu \nu }^{\left(
3\right) }+\frac{n\left( n+1\right) }{6}\Lambda g_{\mu \nu }=T_{\mu \nu },
\end{equation}%
where 
\begin{gather}
G_{\mu \nu }^{\left( 3\right) }=-3\left( 4R_{\qquad }^{\tau \rho \sigma
\kappa }R_{\sigma \kappa \lambda \rho }R_{~\nu \tau \mu }^{\lambda
}-8R_{\quad \lambda \sigma }^{\tau \rho }R_{\quad \tau \mu }^{\sigma \kappa
}R_{~\nu \rho \kappa }^{\lambda }\right. \\
+2R_{\nu }^{\ \tau \sigma \kappa }R_{\sigma \kappa \lambda \rho }R_{\quad
\tau \mu }^{\lambda \rho }-R_{\qquad }^{\tau \rho \sigma \kappa }R_{\sigma
\kappa \tau \rho }R_{\nu \mu }+8R_{\ \nu \sigma \rho }^{\tau }R_{\quad \tau
\mu }^{\sigma \kappa }R_{\ \kappa }^{\rho }  \notag \\
+8R_{\ \nu \tau \kappa }^{\sigma }R_{\quad \sigma \mu }^{\tau \rho }R_{\
\rho }^{\kappa }+4R_{\nu }^{\ \tau \sigma \kappa }R_{\sigma \kappa \mu \rho
}R_{\ \tau }^{\rho }-4R_{\nu }^{\ \tau \sigma \kappa }R_{\sigma \kappa \tau
\rho }R_{\ \mu }^{\rho }  \notag \\
+4R_{\qquad }^{\tau \rho \sigma \kappa }R_{\sigma \kappa \tau \mu }R_{\nu
\rho }+2RR_{\nu }^{\ \kappa \tau \rho }R_{\tau \rho \kappa \mu }+8R_{\ \nu
\mu \rho }^{\tau }R_{\ \sigma }^{\rho }R_{\ \tau }^{\sigma }  \notag \\
-8R_{\ \nu \tau \rho }^{\sigma }R_{\ \sigma }^{\tau }R_{\ \mu }^{\rho
}-8R_{\quad \sigma \mu }^{\tau \rho }R_{\ \tau }^{\sigma }R_{\nu \rho
}-4RR_{\ \nu \mu \rho }^{\tau }R_{\ \tau }^{\rho }  \notag \\
+4R_{\quad }^{\tau \rho }R_{\rho \tau }R_{\nu \mu }-8R_{\ \nu }^{\tau
}R_{\tau \rho }R_{\ \mu }^{\rho }+4RR_{\nu \rho }R_{\ \mu }^{\rho }  \notag
\\
\left. -R^{2}R_{\nu \mu }\right) -\frac{1}{2}\mathcal{L}_{\left( 3\right)
}g_{\mu \nu }.  \notag
\end{gather}%
$.$

\subsection{EPYMGBL Black hole solution for $N\left( =n+2\right) \geq 7$
dimensions}

As before we start with the $rr$ component of Einstein equation which reads%
\begin{gather}
\frac{3\left( n\left( n-1\right) Q^{2}\right) ^{q}}{r^{4\left( q-1\right) }}%
+3n\left[ \left( r^{5}-2\tilde{\alpha}_{2}r^{3}g\left( r\right)
+3rg^{2}\right) g^{\prime }\left( r\right) +\tilde{\alpha}_{3}\left(
n-5\right) r^{2}g\left( r\right) ^{3}\right. -  \notag \\
\left. \tilde{\alpha}_{2}\left( n-3\right) r^{2}g\left( r\right)
^{2}+r^{4}\left( n-1\right) g\left( r\right) +\frac{\Lambda }{3}\left(
n+1\right) r^{6}\right] =0,
\end{gather}%
where $\widetilde{\alpha }_{3}=\left( n-1\right) \left( n-2\right) \left(
n-3\right) \left( n-4\right) \alpha _{3}.$

\subsubsection{The particular case of $\tilde{\protect\alpha}_{3}=\tilde{%
\protect\alpha}_{2}^{2}/3$}

In the third order Lovelock theory we first prefer to impose a condition on
Lovelock's parameters such as $\tilde{\alpha}_{3}=\tilde{\alpha}_{2}^{2}/3.$
This helps us to work with less complicity and in the sequel for the sake of
completeness we shall present the general solution without this restriction
as well. The metric function after this condition is given by%
\begin{equation}
f\left( r\right) =\left\{ 
\begin{tabular}{ll}
$1+\frac{r^{2}}{\tilde{\alpha}_{2}}\left( 1-\sqrt[3]{1+\Lambda \tilde{\alpha}%
_{2}+\frac{12m\tilde{\alpha}_{2}}{nr^{n+1}}+\frac{3\tilde{\alpha}_{2}Q_{1}}{%
r^{4q}}}\right) ,$ & $q\neq \frac{n+1}{4}$ \\ 
$1+\frac{r^{2}}{\tilde{\alpha}_{2}}\left( 1-\sqrt[3]{1+\Lambda \tilde{\alpha}%
_{2}+\frac{12m\tilde{\alpha}_{2}}{nr^{n+1}}+\frac{3\tilde{\alpha}%
_{2}Q_{2}\ln r}{r^{n+1}}}\right) ,$ & $q=\frac{n+1}{4}$%
\end{tabular}%
\right.
\end{equation}%
where as usual $m$ is the mass of the black hole. One may find 
\begin{equation}
\lim_{r\rightarrow \infty }f\left( r\right) \rightarrow 1+\frac{r^{2}}{%
\tilde{\alpha}_{2}}\left( 1-\sqrt[3]{1+\Lambda \tilde{\alpha}_{2}}\right) ,%
\text{ \ \ \ \ \ }\left( \Lambda >0\right)
\end{equation}%
which gives the asymptotical behavior of the metric such as de Sitter, Anti
de Sitter or flat $\left( \Lambda =0\right) $. We note that in the limit $%
\tilde{\alpha}_{2}\rightarrow 0,$ we have $f\left( r\right) \rightarrow 1-%
\frac{\Lambda }{3}r^{2},$ as it should.

\subsubsection{The case of arbitrary $\tilde{\protect\alpha}_{2},\tilde{%
\protect\alpha}_{3}$}

The general solution of the metric function for the case of EPMGBL is given
by%
\begin{equation}
f\left( r\right) =\left\{ 
\begin{tabular}{ll}
$1+\frac{\tilde{\alpha}_{2}r^{2}}{3\tilde{\alpha}_{3}}\left( 1+\frac{\sqrt[3]%
{\Delta }}{2\omega n\tilde{\alpha}_{2}r^{n+1+2q}}+\frac{2\omega n\left( 
\tilde{\alpha}_{2}^{2}-3\tilde{\alpha}_{3}\right) r^{n+1+2q}}{\sqrt[3]{%
\Delta }\tilde{\alpha}_{2}}\right) ,$ & $q\neq \frac{n+1}{4}$ \\ 
$1+\frac{\tilde{\alpha}_{2}r^{2}}{3\tilde{\alpha}_{3}}\left( 1+\frac{\sqrt[3]%
{\tilde{\Delta}}}{2n\tilde{\alpha}_{2}r^{n+1}}+\frac{2n\left( \tilde{\alpha}%
_{2}^{2}-3\tilde{\alpha}_{3}\right) r^{n+1}}{\sqrt[3]{\tilde{\Delta}}\tilde{%
\alpha}_{2}}\right) ,$ & $q=\frac{n+1}{4}$%
\end{tabular}%
\right.
\end{equation}

where%
\begin{eqnarray}
\Delta &=&36\omega ^{2}n^{2}r^{2\left( 1+n+q\right) }\left( \sqrt{\delta }%
\tilde{\alpha}_{3}-3Q_{1}^{2}\tilde{\alpha}_{3}^{2}r^{1+n}-\omega
r^{4q}\zeta \right) , \\
\delta &=&\left( 3Q_{1}\tilde{\alpha}_{3}^{2}r^{1+n}\right)
^{2}+6Q_{1}\omega r^{4q+1+n}\zeta +\frac{\omega ^{2}r^{8q}}{\tilde{\alpha}%
_{3}^{2}}\left\{ \zeta ^{2}-\left( \tilde{\alpha}_{2}^{2}-3\tilde{\alpha}%
_{3}\right) ^{3}\left( \frac{2n}{9}r^{1+n}\right) ^{2}\right\} ,  \notag \\
\zeta &=&\lambda nr^{1+n}+\tilde{\alpha}_{3}^{2}m,\text{ \ \ }\omega =1+n-4q,%
\text{ \ \ }  \notag \\
\lambda &=&\alpha _{3}^{2}\Lambda +\tilde{\alpha}_{2}\tilde{\alpha}_{3}-%
\frac{2}{9}\tilde{\alpha}_{2}^{3},\text{ \ \ }Q_{1}=n^{q}\left( n-1\right)
^{q}Q^{2q},  \notag
\end{eqnarray}%
and 
\begin{eqnarray}
\tilde{\Delta} &=&36n^{2}r^{2\left( 1+n\right) }\left( 3\tilde{\alpha}_{3}%
\sqrt{\tilde{\delta}}-\tilde{\zeta}\right) , \\
\tilde{\delta} &=&-\left( \frac{2nr^{1+n}}{3\tilde{\alpha}_{3}}\right)
^{2}\left( \tilde{\alpha}_{2}^{2}-3\tilde{\alpha}_{3}\right) ^{3}+\frac{9}{%
\tilde{\alpha}_{3}^{2}}\tilde{\zeta}^{2},  \notag \\
\tilde{\zeta} &=&\tilde{\alpha}_{3}^{2}\chi +\lambda nr^{1+n},\text{ \ \ }%
\chi =3Q_{2}\ln r+m,\text{ \ \ }Q_{2}=n^{\frac{1+n}{4}}\left( n-1\right) ^{%
\frac{1+n}{4}}Q^{\frac{1+n}{2}}.  \notag
\end{eqnarray}%
Occurrence of the roots naturally restricts the ranges of parameters since
the results must be real and physically admissible.

Here also one can find the nature of metric at infinity, namely%
\begin{equation}
\lim_{r\rightarrow \infty }f\left( r\right) \rightarrow 1+\Lambda _{eff}r^{2}
\end{equation}%
where%
\begin{equation}
\Lambda _{eff}=\frac{1}{9\tilde{\alpha}_{3}}\left( -9\sqrt[3]{\frac{\lambda 
}{6}}+\left( 3\tilde{\alpha}_{3}-\tilde{\alpha}_{2}^{2}\right) \sqrt[3]{%
\frac{6}{\lambda }}+3\tilde{\alpha}_{2}\right) .
\end{equation}

\subsection{Thermodynamics of the EPYMGBL black hole}

As before, we complete this chapter by giving some thermodynamical
properties of the EPYMGB black hole solution. Clearly, working analytically
with the arbitrary $\tilde{\alpha}_{2},\tilde{\alpha}_{3}$ may not be
possible therefore we only stress on the specific case of $\tilde{\alpha}%
_{3}=\tilde{\alpha}_{2}^{2}/3.$ Given this particular choice, we start with
the ADM mass of the BH which reads%
\begin{equation}
\check{M}_{ADM}=\check{m}=\left\{ 
\begin{tabular}{ll}
$\frac{n}{4}\left( 1+\check{\alpha}_{2}\left( \frac{\check{\alpha}_{2}}{3}%
+1\right) -\check{Q}_{1}-\frac{\check{\Lambda}}{3}\right) $ & $q\neq \frac{%
n+1}{4}$ \\ 
$\frac{n}{4}\left( 1+\check{\alpha}_{2}\left( \frac{\check{\alpha}_{2}}{3}%
+1\right) -\check{Q}_{1}\ln \left( r_{h}\right) -\frac{\check{\Lambda}}{3}%
\right) $ & $q=\frac{n+1}{4}$%
\end{tabular}%
\right. ,
\end{equation}%
whose Hawking temperature is given by%
\begin{equation}
\check{T}_{H}=\left\{ 
\begin{tabular}{ll}
$\frac{-\check{Q}_{1}\left( n+1-4q\right) -\frac{\check{\Lambda}}{3}\left(
n+1\right) +\left( n-5\right) \frac{\check{\alpha}_{2}^{2}}{3}+\check{\alpha}%
_{2}\left( n-3\right) +n-1}{4\pi \left( 1+\check{\alpha}_{2}\right) ^{2}}$ & 
$q\neq \frac{n+1}{4}$ \\ 
$\frac{-\check{Q}_{1}-\frac{\check{\Lambda}}{3}\left( n+1\right) +\left(
n-5\right) \frac{\check{\alpha}_{2}^{2}}{3}+\check{\alpha}_{2}\left(
n-3\right) +n-1}{4\pi \left( 1+\check{\alpha}_{2}\right) ^{2}}$ & $q=\frac{%
n+1}{4}$%
\end{tabular}%
\right. .
\end{equation}%
We notice here that the Hawking temperature diverges as $\check{\alpha}_{2}$
approaches to $-1.$

\subsubsection{$q\neq \frac{n+1}{4}$}

The heat capacity:%
\begin{equation}
\begin{tabular}{l}
$\check{C}=\frac{n}{2}\pi ^{\frac{n+1}{2}}\frac{\left( \check{\alpha}%
_{2}+1\right) }{\Gamma \left( \frac{n+1}{2}\right) }\frac{\frac{\check{%
\Lambda}}{3}\left( n+1\right) +\left( n+1-4q\right) \check{Q}_{1}-\left(
n-5\right) \frac{\check{\alpha}_{2}^{2}}{3}-\left( n-3\right) \check{\alpha}%
_{2}-\left( n-1\right) }{\left\{ \left( 1+n\right) \left( 5\check{\alpha}%
_{2}+1\right) \frac{\check{\Lambda}}{3}-4\left[ \left( q-\frac{5}{4}\right) 
\check{\alpha}_{2}+\left( q-\frac{1}{4}\right) \right] \check{Q}_{1}\left(
1+n-4q\right) +\left( n-5\right) \frac{\check{\alpha}_{2}^{3}}{3}+\frac{2}{3}%
\left( n-8\right) \check{\alpha}_{2}^{2}-6\check{\alpha}_{2}+n-1\right\} },$%
\end{tabular}%
\end{equation}

The free energy:%
\begin{equation}
\begin{tabular}{l}
$\check{F}=\frac{\left\{ \left[ \check{Q}_{1}\left( n+1-4q\right) +\frac{%
\check{\Lambda}\left( n+1\right) }{3}-\left( n-5\right) \frac{\check{\alpha}%
_{2}^{2}}{3}-\left( n-3\right) \check{\alpha}_{2}+1-n\right] \pi ^{\frac{n-1%
}{2}}-2n\left( \check{Q}_{1}+\frac{\check{\Lambda}}{3}-1-\check{\alpha}_{2}-%
\frac{\check{\alpha}_{2}^{2}}{3}\right) \left( \check{\alpha}_{2}+1\right)
^{2}\Gamma \left( \frac{n+1}{2}\right) \right\} }{8\left( 1+\check{\alpha}%
_{2}\right) ^{2}\Gamma \left( \frac{n+1}{2}\right) }.$%
\end{tabular}%
\end{equation}

\subsubsection{$q=\frac{n+1}{4}$}

The heat capacity:%
\begin{equation}
\begin{tabular}{l}
$\check{C}=\frac{n}{2}\pi ^{\frac{n+1}{2}}\frac{\left( \check{\alpha}%
_{2}+1\right) }{\Gamma \left( \frac{n+1}{2}\right) }\frac{\frac{\check{%
\Lambda}}{3}\left( n+1\right) +\check{Q}_{2}-\left( n-5\right) \frac{\check{%
\alpha}_{2}^{2}}{3}-\left( n-3\right) \check{\alpha}_{2}-\left( n-1\right) }{%
\left\{ \left( 1+n\right) \left( 5\check{\alpha}_{2}+1\right) \frac{\check{%
\Lambda}}{3}-\left[ \left( n-4\right) \check{\alpha}_{2}+n\right] \check{Q}%
_{2}+\left( n-5\right) \frac{\check{\alpha}_{2}^{3}}{3}+\frac{2}{3}\left(
n-8\right) \check{\alpha}_{2}^{2}-6\check{\alpha}_{2}+n-1\right\} },$%
\end{tabular}%
\end{equation}

The free energy:%
\begin{equation}
\begin{tabular}{l}
$\check{F}=\frac{\left[ \check{Q}_{2}+\frac{\check{\Lambda}\left( n+1\right) 
}{3}-\left( n-5\right) \frac{\check{\alpha}_{2}^{2}}{3}-\left( n-3\right) 
\check{\alpha}_{2}+1-n\right] \pi ^{\frac{n-1}{2}}-2n\left( \check{Q}_{2}\ln
\left( r_{h}\right) +\frac{\check{\Lambda}}{3}-1-\check{\alpha}_{2}-\frac{%
\check{\alpha}_{2}^{2}}{3}\right) \left( \check{\alpha}_{2}+1\right)
^{2}\Gamma \left( \frac{n+1}{2}\right) }{8\left( 1+\check{\alpha}_{2}\right)
^{2}\Gamma \left( \frac{n+1}{2}\right) }.$%
\end{tabular}%
\end{equation}%
In the foregoing expressions it is observed that for $\check{\alpha}_{2}=-1,$
the free energy diverges, signalling the occurrence of a critical point.
Further, the sign of the heat capacity can be investigated to see whether
thermodynamically the system is stable ($\check{C}>0$) or unstable ($\check{C%
}<0$), which will be ignored in this Letter.

\section{Yang-Mills hierarchies}

In this section we investigate the possible black hole solutions for the
case of a superposition of the different power of the YM invariant $\mathcal{%
F}$ and any further investigation in this line is going to be part of our
future study. It is our belief that a detailed analysis of the energy
conditions for the YM hierarchy exceeds the limitations of the present
Letter, we shall therefore ignore it. The YM hierarchies in $d-$dimensions
has been studied by D. H. Tchrakian, et. al. \cite{10} in a different sense.
Here we start with an action in the form of 
\begin{equation}
I=\frac{1}{2}\int_{\mathcal{M}}d^{n+2}x\sqrt{-g}\left( R+\alpha _{2}\mathcal{%
L}_{GB}+\alpha _{3}\mathcal{L}_{(3)}-\overset{q}{\underset{k=0}{\tsum }b_{k}}%
\mathcal{F}^{k}\right) ,
\end{equation}%
in which $\mathcal{F}$ is the YM invariant, $b_{0}=\frac{n\left( n+1\right) 
}{3}\Lambda $ and $b_{k\geq 1}$ is a coupling constant. Variation of the
action with respect to the spacetime metric $g_{\mu \nu }$ yields the field
equations%
\begin{gather}
G_{\ \nu }^{\mu }+\alpha _{2}G_{\mu \nu }^{\nu GB}+\alpha _{3}G_{\mu }^{\nu
\left( 3\right) }=T_{\ \nu }^{\mu }, \\
T_{\ \nu }^{\mu }=-\frac{1}{2}\overset{q}{\underset{k=0}{\tsum }b_{k}}\left(
\delta _{\ \nu }^{\mu }\mathcal{F}^{k}-4k\mathbf{Tr}\left( F_{\nu \lambda
}^{\left( a\right) }F^{\left( a\right) \ \mu \lambda }\right) \mathcal{F}%
^{k-1}\right) ,
\end{gather}%
and variation with respect to the gauge potentials $\mathbf{A}^{\left(
a\right) }$ yields the YM equations%
\begin{equation}
\overset{q}{\underset{k=0}{\tsum }b_{k}}\left\{ \mathbf{d}\left( ^{\star }%
\mathbf{F}^{\left( a\right) }\mathcal{F}^{k-1}\right) +\frac{1}{\sigma }%
C_{\left( b\right) \left( c\right) }^{\left( a\right) }\mathcal{F}^{k-1}%
\mathbf{A}^{\left( b\right) }\wedge ^{\star }\mathbf{F}^{\left( c\right)
}\right\} =0.
\end{equation}%
Our metric ansatz for $N\left( =n+2\right) $ dimensions, is given by (7) and
the YM field ansatz is as before such that the new energy momentum tensor
reads as 
\begin{equation}
T_{\text{ }b}^{a}=-\frac{1}{2}\overset{q}{\underset{k=0}{\tsum }b_{k}}%
\mathcal{F}^{k}\text{diag}\left[ 1,1,\gamma ,\gamma ,..,\gamma \right] ,%
\text{ \ and \ }\gamma =\left( 1-\frac{4k}{n}\right) .
\end{equation}%
The solution of Einstein equation for $\alpha _{2}=\alpha _{3}=0$ reveals
that%
\begin{equation}
f\left( r\right) =1-\frac{4m}{nr^{n-1}}-\frac{1}{n}\Psi 
\end{equation}%
where $m$ is the ADM mass of the black hole and 
\begin{eqnarray}
\Psi  &=&\tint r^{n}\overset{q}{\underset{k=0}{\tsum }b_{k}}\mathcal{F}%
^{k}dr=  \notag \\
&&\left\{ 
\begin{tabular}{ll}
$\overset{q}{\underset{k=0}{\tsum }b_{k}}\frac{\left( n\left( n-1\right)
Q^{2}\right) ^{k}}{\left( n-4k+1\right) r^{4k-n-1}},$ & $k\neq \frac{n+1}{4}$
\\ 
$b_{\frac{n+1}{4}}\left( n\left( n-1\right) Q^{2}\right) ^{\frac{n+1}{4}}\ln
r+\overset{q}{\underset{k=0\neq \frac{n+1}{4}}{\tsum }b_{k}}\frac{\left(
n\left( n-1\right) Q^{2}\right) ^{k}}{\left( n-4k+1\right) r^{4k-n-1}},$ & $%
k^{\prime }=\frac{n+1}{4}$%
\end{tabular}%
\right. .
\end{eqnarray}%
The case of GB which comes after $\alpha _{3}=0$ reveals%
\begin{equation}
f_{\pm }\left( r\right) =1+\frac{r^{2}}{2\tilde{\alpha}_{2}}\left( 1\pm 
\sqrt{1+\frac{16m\tilde{\alpha}_{2}}{nr^{n+1}}+\frac{4\tilde{\alpha}_{2}}{%
nr^{n+1}}\Psi }\right) .
\end{equation}%
For the case of $\alpha _{2},\alpha _{3}\neq 0$ first we give a solution for
the specific choice of $\tilde{\alpha}_{3}=\tilde{\alpha}_{2}^{2}/3$ which
admits%
\begin{equation}
f\left( r\right) =1+\frac{r^{2}}{\tilde{\alpha}_{2}}\left( 1-\sqrt[3]{1+%
\frac{12m\tilde{\alpha}_{2}}{nr^{n+1}}+\frac{3\tilde{\alpha}_{2}}{nr^{n+1}}%
\Psi }\right) ,
\end{equation}%
and then the most general solution for $\alpha _{2},\alpha _{3}\neq 0$
yields a general metric function as%
\begin{equation}
f\left( r\right) =1+\frac{\tilde{\alpha}_{2}r^{2}}{3\tilde{\alpha}_{3}}%
\left( 1+\frac{\sqrt[3]{\Delta }}{2n\tilde{\alpha}_{2}r^{n+1}}+\frac{%
2n\left( \tilde{\alpha}_{2}^{2}-3\tilde{\alpha}_{3}\right) r^{n+1}}{\sqrt[3]{%
\Delta }\tilde{\alpha}_{2}}\right) ,
\end{equation}%
where%
\begin{equation}
\Delta =36n^{2}r^{2\left( 1+n\right) }\left( \frac{\tilde{\alpha}_{3}}{3}%
\sqrt{3}\sqrt{\delta }+\left( \tilde{\alpha}_{3}-\frac{2}{9}\tilde{\alpha}%
_{2}^{2}\right) -6\tilde{\alpha}_{3}^{2}\left( \frac{1}{2}\Psi +m\right)
\right) ,
\end{equation}%
and%
\begin{equation}
\delta =\left( 4\tilde{\alpha}_{3}-\tilde{\alpha}_{2}^{2}\right)
n^{2}r^{2\left( n+1\right) }+36\tilde{\alpha}_{2}nr^{n+1}\left( \tilde{\alpha%
}_{3}-\frac{2}{9}\tilde{\alpha}_{2}^{2}\right) \left( \frac{1}{2}\Psi
+m\right) +108\tilde{\alpha}_{3}^{2}\left( \frac{1}{2}\Psi +m\right) ^{2}.
\end{equation}

\section{Conclusion}

Clearly, the YM invariant /source $\mathcal{F}^{q}$ becomes simplest for $q=1
$. Beside simplicity there is no valid argument that prevents us from
choosing $q\neq 1.$ As a result, the latter modifies many black holes
obtained from YM field as a source and it specifies also in higher
dimensions, which $q$ values are consistent with the energy conditions. For
electric type fields there is a drawback that $\mathcal{F}^{q}$ may not be
real for any $q$, however, this doesn't arise for our pure magnetic type YM
field. We note that the same situation is valid also in the power-Maxwell
case. In spite of so much non-linearities, including even a YM source of the
form $\overset{q}{\underset{k=0}{\tsum }b_{k}}\mathcal{F}^{k},$\ with the
requirement of spherical symmetry we obtained exact black hole solutions to
the Lovelock's third order theory. In analogy with the non-linear
electrodynamics, the requirement of conformal invariance puts further
restrictions on $q$ and the spacetime, namely the dimension of spacetime
turns out to be a multiple of 4. Physically, the power $q$ modifies the
strength of fields both for $r\rightarrow 0$ and $r\rightarrow \infty $. It
is observed that asymptotically $\left( r\rightarrow \infty \right) ,$
irrespective of $q$ the effect of Lovelock gravity, whether at second or
third order, becomes equivalent to an effective cosmological constant.

\bigskip

\textbf{Tables:}%
\begin{equation}
\begin{tabular}{|l|l|l|l|l|}
\hline
& WEC & SEC & DEC & CC \\ \hline
$q<0$ & no & no & no & no \\ \hline
$0\leq q<\frac{n}{4}$ & yes & no & no & no \\ \hline
$\frac{n}{4}\leq q<\frac{n+1}{4}$ & yes & yes & no & no \\ \hline
$\frac{n+1}{4}\leq q<\frac{n+1}{2}$ & yes & yes & yes & yes \\ \hline
$\frac{n+1}{2}<q$ & yes & yes & yes & no \\ \hline
\end{tabular}%
.  \tag{Tab. (1)}
\end{equation}

\begin{equation}
\begin{tabular}{|l|l|l|l|l|l|l|l|l|l|}
\hline
dimensions $N$ & $\mathbf{5}$ & $6$ & $7$ & $8$ & $\mathbf{9}$ & $10$ & $11$
& $12$ & $\mathbf{13}$ \\ \hline
possible integer $q$ & $\mathbf{1}$ & $2$ & $2$ & $2,3$ & $\mathbf{2},3$ & $%
3,4$ & $3,4$ & $3,4,5$ & $\mathbf{3},4,5$ \\ \hline
\end{tabular}%
.  \tag{Tab. (2)}
\end{equation}

\textbf{Table Captions:}

Tab. 1: Energy conditions WEC, SEC and DEC and the causality condition (CC)
versus the admissible ranges of parameter $q$.

Tab. 2: List of some possible integer $q$ values versus $N$.

\end{document}